\documentclass{article}
\usepackage{procla}
\begin{document}
\centerline{\normalsize\bf THE MIXING OF QUARKS AND LEPTONS}
\baselineskip=22pt
\centerline{\normalsize\bf AND NEUTRINO DEGENERACY
\footnote{Talk given by one of us (H.F.) at the Ringberg 
Euroconference on New Trends in Neutrino Physics,
Ringberg, Germany, May 1998.}}
\baselineskip=16pt
%\centerline{\normalsize\bf MANUSCRIPT BY COMPUTER}
%\centerline{\footnotesize\sf (For subsequent 20\% photoreduction
%to 17.8 $\times$ 11.9 cm text area)\footnote{The \LaTeX\ source
%file for this document may be used as a template for your
%article, and can be requested by e-mailing {\sf
%worldscp@singnet.com.sg}.}}
%\vfill
\vspace*{0.6cm}
\centerline{\footnotesize HARALD FRITZSCH}
\baselineskip=13pt
\centerline{\footnotesize\it Sektion Physik, Universit\"at M\"unchen,
Theresienstrasse 37, 80333 M\"unchen, Germany}
\baselineskip=12pt
%\centerline{\footnotesize\it City, State ZIP/Zone, Country}
\centerline{\footnotesize E-mail: bm@hep.physik.uni-muenchen.de}
\vspace*{0.3cm}
\centerline{\footnotesize and}
\vspace*{0.3cm}
\centerline{\footnotesize ZHI-ZHONG XING}
\baselineskip=13pt
\centerline{\footnotesize\it Sektion Physik, Universit\"at M\"unchen,
Theresienstrasse 37, 80333 M\"unchen, Germany}
\centerline{\footnotesize E-mail: xing@hep.physik.uni-muenchen.de}

%\vfill
\vspace*{0.9cm}
\abstracts{In drawing on an analogy with the flavor mixing 
observed in the quark sector we discuss a pattern of large 
flavor mixing angles in the lepton sector. Simple arguments
based on a democratic symmetry and its violation in the 
lepton sector allow us to determine the flavor mixing matrix
of leptons. The mixing angle relevant for solar neutrino
oscillations is maximal (close to $45^{\circ}$), while the
angle relevant for atmospheric neutrino oscillations is
given by $\sin^2 2 \theta = 8/9$. The emerging pattern
is consistent with the results of the solar and
atmospheric neutrino experiments.}
\vspace*{0.6cm}
\normalsize\baselineskip=15pt
\setcounter{footnote}{0}
\renewcommand{\thefootnote}{\alph{footnote}}

\section{Main Chapter}
In this talk I shall concentrate on phenomenological issues of the
neutrino mixing phenomenon and in particular on analogies between the leptonic
mixing and the mixing of the quark flavors. The mixing of quark
flavors is known for almost forty years, and there is little doubt that there
will be parallelisms between the mixings of leptonic and quark flavors.
Nevertheless, substantial differences might exist.

Let me first discuss some general features of the quark and charged
lepton mass spectra, and of flavor mixing.

I should like to
emphasize, that the term ``neutrino mixing'' which is often used is
misleading. If neutrino
oscillations exist, they manifest a general leptonic mixing phenomenon and a
mismatch between the neutrino mass spectrum and the mass spectrum of the
charged leptons, in analogy to the quarks. Just as the flavor
mixing angles in the quark sector are related intrinsically to the quark
masses, the parameters of the neutrino oscillations or in general the leptonic
mixing angles are related directly both to the neutrino and the charged
lepton mass terms. In particular the pattern of the charged lepton masses
will be significant for the leptonic flavor mixing and for the magnitude of
the neutrino oscillations.

The mass spectra of the quarks and of the charged leptons are similar. Both
for the quarks and for the charged leptons the spectra are largely dominated
by the members of the third family. 95\% of the lepton masses are provided
by the $\tau$--lepton. The $b$--quark contributes about 97\% to the masses in
the charge $-1/3$ sector. The charge $+2/3$ sector is dominated to 98\% by
the $t$--quark. Both in case of the charged leptons and of the quarks the
contribution of the first family to the total mass in the corresponding
channel is almost negligible. At the
present time it is not known whether such a conspicuous hierarchy in the mass
spectra of the charged fermions is accompanied by a similar hierarchy of the
neutrinos. If such a hierarchy would also exist in the neutrino sector, one
would expect that the $\tau$--neutrino is the heaviest neutrino, accompanied
by a relatively light $\mu$--neutrino and an almost massless
$e$--neutrino.

For the discussion of flavor mixing it is often useful to treat the fermion
masses as parameters, which can be changed arbitrarily
and in particular set to zero or infinity.
Obviously the physics of the leptons and quarks will not be changed
significantly, if we set the masses of the first and second family to zero.
The departure from the real world will be about 5\%. Of course, due to our
ignorance about the origin of the fermion masses we do not know whether such
a change of these masses is indeed possible. Within the framework of the
standard model it is, of course, easy to make such changes just by modifying
the coupling parameters describing the interaction of the fermions and the
scalar field.

If the masses of the first two families vanish, the mass matrices of the
fermions become matrices of rank one:
\begin{equation}
M = C \left( \begin{array}{ccc}
		0 & 0 & 0 \\
		0 & 0 & 0 \\
		0 & 0 & 1
		\end{array} \right) \, .
%               (1)
\end{equation}
In the limit there appears to be a mass gap $C$, given by the mass of the
$t$--quark, the $b$--quark or the $\tau $--lepton respectively. By a suitable
orthogonal transformation the mass matrix $M$
can be brought into a form, in which
all matrix elements are identical:
\begin{equation}
\overline{M} = \frac{C}{3} \left( \begin{array}{ccc} 
				   1 & 1 & 1\\
				   1 & 1 & 1\\
				   1 & 1 & 1
				   \end{array} \right) \, .
%               (2)
\end{equation}
Such a mass matrix, which is often called a democratic mass matrix, also
has rank one. As far as the charged leptons and the quarks are concerned, we
can speak either of a hierarchy basis ($M$)
or of a democratic basis ($\overline{M}$). Both are,
of course, equivalent. However, in the democratic basis one realizes a new
symmetry described by the discrete group $S(3)_{\rm L} \times S(3)_{\rm R}$.

Before coming back to this symmetry, let me describe how such
a situation could arise. If we look at the quarks or charged leptons using the
democratic basis one finds that there are universal transitions between all
three fermion states. One is reminded of the so--called ``pairing force'',
which is responsible for the appearance of a mass gap in
superconductivity or for giant resonances in nuclear physics. I should also
like to mention that the mass spectrum of the neutral pseudoscalar mesons in
QCD ($\pi^0, \eta, \eta'$) is described by democratic--type mass matrix.
Between the various $\bar qq$--states there are large
transition elements provided by the gluonic interaction. These transitions
are universal in the chiral limit due to the universality of the
gluonic interaction. In the pseudoscalar channel these transitions are
particularly strong and lead to large mixing effects, due to large
non--perturbative effects. It is due to these gluonic transitions 
that in the limit of chiral $SU(3)_{\rm L} \times SU(3)_{\rm R}$ the
$0^{-+}$--mesons segregate into a massive singlet and a massless
octet.

In the case of the pseudoscalar mesons we can also
denote the transformation between the hierarchy basis and the democratic basis.
The eigenstates of the $S(3)$--symmetry are nothing but the states
$\bar uu$, $\bar dd$ and $\bar ss$. The connection between the mass
eigenstates and the $\bar q q$--states is given by:
\begin{eqnarray}
\pi^0 & = & \frac{1}{\sqrt{2}} \left( \bar uu - \bar dd \right) \; ,
\nonumber \\
\eta  & = & \frac{1}{\sqrt{6}} \left( \bar uu + \bar dd - 2 \bar ss \right)
\; , \nonumber \\
\eta ' & = & \frac{1}{\sqrt{3}} \left( \bar uu + \bar dd + \bar ss
\right) \, .
%               (3)
\end{eqnarray}
In analogy let me write down the mass eigenstates of charged leptons
in the democratic limit in terms of the eigenstates $l_1, l_2$ and $l_3$ of
the democratic symmetry \cite{FH}:
\begin{eqnarray}
e & = & \frac{1}{\sqrt{2}} \left( l_1 - l_2 \right) \; , \nonumber \\
\mu & = & \frac{1}{\sqrt{6}} \left( l_1 + l_2 - 2l_3 \right) \; , \nonumber \\
\tau & = & \frac{1}{\sqrt{3}} \left( l_1 + l_2 + l_3 \right) \, .
%               (4)
\end{eqnarray}
Similar relations can be written down for the quarks. In reality the
democratic symmetry is not exact, but broken by small terms. These symmetry
breaking effects have been discussed some time ago by a number of authors,
and we shall refer to the literature \cite{FH,Dem}.

It is interesting to discuss the description of the flavor mixing in the
context of the democratic symmetry and its violation. In the limit of the
democratic symmetry one expects for the quarks that no flavor mixing is
present. In other words, the flavor mixing angles must be related to the
violation of the symmetry and in particular to the masses of the first two
families, or rather to the mass ratios of the masses of the
first two families and the mass of the corresponding representative of the
third family. The best way
of describing the flavor mixing would be one in which the parameters for the
flavor mixing, e. g. the flavor mixing angles, are smooth functions of the
symmetry breaking parameters. In view of this we have recently studied all
possible ways for describing the flavor mixing of the quarks. As discussed
in Ref. \cite{FX98} 
there exist nine different ways in general to describe the
mixing of three families. But only one description obeys the constraints
discussed above. In particular the so-called "standard" 
parametrization advocated
by Particle Data Group \cite{PDG}
does not obey these constraints. Instead one is lead to the following
parametrization \cite{FX97}:
\begin{eqnarray}
V & = & \left( \matrix{
	   c_{\rm u} & s_{\rm u} & 0 \cr
	  -s_{\rm u} & c_{\rm u} & 0 \cr
	   0   & 0   & 1 \cr } \right)  
					\left( \matrix{
					e^{- i \varphi} &  0 & 0 \cr
					0               &  c & s \cr
					0               & -s & c \cr } \right )
	  \left(  \matrix{
		   c_{\rm d} & -s_{\rm d} & 0 \cr
		   s_{\rm d} & c_{\rm d} & 0 \cr
		   0   & 0   & 1 \cr } \right) \nonumber \\ \nonumber \\
& = & \left( \matrix{
	s_{\rm u} \, s_{\rm d} \, c + c_{\rm u} \, c_{\rm d} \, 
	e^{-i \varphi} & s_{\rm u} \, c_{\rm d} \, c -  
	c_{\rm u} \, s_{\rm d} \, e^{-i \varphi} & s_{\rm u} s \cr
	c_{\rm u} \, s_{\rm d} \, c - s_{\rm u} \, c_{\rm d} \, 
	e^{-i \varphi} & c_{\rm u} \, c_{\rm d} \, c +
	s_{\rm u} \, \, s_{\rm d} \, e^{-i \varphi} & c_{\rm u} \, s \cr
	- s_{\rm d} \, s & - c_{\rm d} \, s & c \cr } \right) \, .
%               (5)
\end{eqnarray}
Here $c_{\rm u} = \cos\theta_{\rm u}$, $s_{\rm d}=\sin \theta_{\rm d}$,
$c =\cos\theta$, etc.

It is interesting to note that this way of describing the flavor mixing matrix
of the quarks is in the absence of the complex phase $\varphi $ identical to
the rotation matrix given originally by Euler \cite{Euler}. 
Since it will turn out that the
corresponding mixing angles are small, one finds in a good approximation:
\begin{equation}
V = \left( \begin{array}{ccc}
	   e^{- i \, \varphi} & s_{\rm u} - s_{\rm d} \, 
	   e^{-i \, \varphi} & s_{\rm u} \, s \\
	   s_{\rm d} - s_{\rm u} \, e^{i \, \varphi} & 1 & s \\ 
	   - s_{\rm d} \, s & -s & 1
	   \end{array} \right) \, .
%               (6)
\end{equation}
As discussed in Ref. \cite{FX97} the three rotation angles
$\theta_{\rm u}$, $\theta_{\rm d}$
and $\theta $ have a precise physical meaning. The angle $\theta $ is a
combined effect arising from the mixing between the second and third family
(heavy quark mixing). The angle $\theta_{\rm u}$ primarily
describes a mixing between $u$ and
$c$ quarks, and the angle $\theta_{\rm d}$ primarily describes
a mixing between $d$ and $s$ quarks. The angle
$\theta $ is essentially given by the magnitude of $V_{cb}$. The angles
$\theta_{\rm u}$ and $\theta_{\rm d}$ are determined as follows:
\begin{eqnarray}
\tan \, \theta_{\rm u} & = & \left| \frac{V_{ub}}{V_{cb}} \right| \; ,
\nonumber \\
\tan \, \theta_{\rm d} & = & \left| \frac{V_{td}}{V_{ts}} \right| \, .
%               (7)
\end{eqnarray}
In a simple model of symmetry breaking (see Refs. \cite{F78,FX99}) 
these two angles are
related in a simple way to the ratio of quark mass eigenvalues:
\begin{eqnarray}
\theta_{\rm u} & = & \arctan \, \sqrt{\frac{m_u}{m_c}} \; , \nonumber \\
\theta_{\rm d} & = & \arctan \, \sqrt{\frac{m_d}{m_s}} \, .
%               (8)
\end{eqnarray}
Recently the angles as well as the complex phase $\varphi $ have been
determined with relatively high accuracy from a global analysis of
current data \cite{Stocchi}. One finds
\begin{eqnarray}
\theta ~ & = & 2.30^{\circ} \pm 0.09^{\circ } \; , \nonumber \\
\theta_{\rm u} & = & 4.87^{\circ} \pm 0.86^{\circ } \; , \nonumber \\
\theta_{\rm d} & = & 11.71^{\circ} \pm 1.09^{\circ} \; , \nonumber \\
\varphi  & = & 91.1^{\circ} \pm 11.8^{\circ } \, .
%               (9)
\end{eqnarray}
We note that the values obtained here are in very good agreement with the
expectations from the quark masses (see Ref. \cite{FX99}).
Furthermore the complex
phase is in very good agreement with the expectation of $90^{\circ}$, as
expected from a very simple symmetry breaking \cite{FX95,Wu}. 
The fact that the phase is
$90^{\circ }$ signifies that $CP$ violation is maximal in the sence
described in Ref. \cite{FX95}. Furthermore the experimental data
support that the mass matrix of the quarks in the hierarchy basis has the
following structure:
\begin{equation}
M (q) = \left( \begin{array}{ccc}
		 0  & a & 0 \\
		 a^* & b' & b \\
		 0  & b & c
		 \end{array} \right) \, .
%               (10)
\end{equation}
The complex phases of the quark mass
matrices can be arranged such that they appear primarily
in the (1,2) and (2,1) matrix elements.

Since the mass spectrum of the charged leptons exhibits a similar hierarchical
pattern as the quarks, it is most natural to suppose that the matrix
structure and the texture properties of the charged lepton mass matrix is
analogous to those of the quark mass matrices. 

The question arises whether the
neutrinos also exhibit a hierarchical mass pattern. It may well be that the
neutrino masses are not hierarchical at all. If they were, we could write the
neutrino mass matrix in analogy to the charged lepton matrix in the
democratic basis as follows:
\begin{equation}
M ( \nu ) = \frac{C_\nu}{3} \left( \begin{array}{ccc}
			  1 & 1 & 1\\
			  1 & 1 & 1\\
			  1 & 1 & 1
			  \end{array} \right) + \Delta M (\nu) \; ,
%               (11)
\end{equation}
where $\Delta M (\nu)$ denotes the perturbative correction.
The constant $C_\nu$ describes the strength of the pairing-force term
in the neutrino channel. The magnitude of this term is related to the mass of
the heaviest neutrino and could be at most about 30 eV, i. e. it must be
about eight orders of magnitude smaller than the charged lepton term
($C_l = m_\tau$). It is
hard to believe that the ratio $C_\nu / C_l$ is simply a tiny number
by accident. It
would be much more natural to suppose that the leading pairing-force term
is completely absent in the neutrino channel. This possibility was discussed
in Ref. \cite{FX96}. The absence of the leading pairing force term 
for the neutrinos would have drastic consequences for the mixing pattern of
the leptons. An interesting possibility is that the eigenstates of the
democratic symmetry for the neutrinos are identical to the mass eigenstates.
Following Ref. \cite{FX96} we write down the following mass matrices
for the charged leptons and the neutrinos:
\begin{eqnarray}
M(l^-) & = & C_l \left( \matrix{
			     1 & 1 & 1 \cr
			     1 & 1 & 1 \cr
			     1 & 1 & 1 \cr } \right )
		+ \left( \matrix{
			\delta_l & 0 & 0 \cr
			0 & \rho_l & 0 \cr
			0 & 0 & \varepsilon_l \cr } \right ) \; , \nonumber \\
M(\nu) & = & {\bf 0} + \left( \matrix{
			   \delta_{\nu} & 0 & 0 \cr
			     0 & \rho_\nu & 0 \cr
			     0 & 0 & \varepsilon_{\nu} \cr } \right ) \; .
%               (12)
\end{eqnarray}
Obviously large mixing phenomena are generated due to the absence of the
pairing-force term for the neutrinos. The electroweak doublets of 
leptons can be written as:
\begin{equation}
\left( \begin{array}{ccc}
	\nu_1 & \nu_2 & \nu_3 \\
	l_1 & l_2 & l_3
       \end{array} \right) \, .
%               (13)
\end{equation}
Here the upper components, the neutrino states, are the mass eigenstates
while their electroweak partners are identical to the democratic eigenstates
$l_i$. We can also perform a unitary transformation and write down the mass
eigenstates for the charged leptons, neglecting the small breaking terms for
the democratic symmetry, and obtain:
\begin{equation}
\left( \begin{array}{ccccc} 
       \frac{1}{\sqrt{2}} \left( \nu_1 - \nu_2 \right) &~~& \frac{1}{\sqrt{6}}
       \left( \nu_1 + \nu_2 - 2 \nu_3 \right) &~~& \frac{1}{\sqrt{3}}
       \left( \nu_1 + \nu_2 + \nu_3 \right) \\
       e^- && \mu^- && \tau^-
	\end{array} \right) \, .
%               (14)
\end{equation}
In analogy to the case of the quarks we describe the leptonic flavor mixing
matrix as follows:
\begin{eqnarray}
V_l & = & \left ( \matrix{
c_\nu   & s_\nu & 0 \cr
-s_\nu  & c_\nu & 0 \cr
0       & 0     & 1 \cr } \right )  \left ( \matrix{
e^{-i\psi}      & 0     & 0 \cr
0       & c     & s \cr
0       & -s    & c \cr } \right )  \left ( \matrix{
c^{~}_l & -s^{~}_l      & 0 \cr
s^{~}_l & c^{~}_l       & 0 \cr
0       & 0     & 1 \cr } \right )  \nonumber \\ \nonumber \\
& = & \left ( \matrix{
s_\nu s^{~}_l c + c_\nu c^{~}_l e^{-i\psi} &
s_\nu c^{~}_l c - c_\nu s^{~}_l e^{-i\psi} &
s_\nu s \cr
c_\nu s^{~}_l c - s_\nu c^{~}_l e^{-i\psi} &
c_\nu c^{~}_l c + s_\nu s^{~}_l e^{-i\psi}   &
c_\nu s \cr 
- s^{~}_l s     & - c^{~}_l s   & c \cr } \right ) \; .
%               (15)
\end{eqnarray}
The leptonic mixing angles are given by $\theta_l$, describing a mixing for
the charged leptons, an angle $\theta $, describing a mixing between the
second and the third family, and an angle $\theta_{\nu}$, describing the
mixing in the neutrino channel. The complex phase, causing $CP$ violation
for the leptons, is denoted by $\psi $. For simplicity 
we assume $CP$ symmetry to be conserved in the leptonic sector.

The electroweak doublets written above give the following leptonic flavor
mixing matrix:
\begin{eqnarray}
V_l & = & \left( \matrix{
	      \frac{1}{\sqrt{2}} & \frac{1}{\sqrt{6}} 
		& \frac{1}{\sqrt{3}} \cr
	     -\frac{1}{\sqrt{2}} & \frac{1}{\sqrt{6}} 
		& \frac{1}{\sqrt{3}} \cr
		0 & -\frac{2}{\sqrt{6}} 
		& \frac{1}{\sqrt{3}} \cr } \right ) \nonumber \\ \nonumber \\
& = & \left ( \matrix{
\frac{1}{\sqrt{2}}      & \frac{1}{\sqrt{2}}    & 0 \cr
-\frac{1}{\sqrt{2}}     & \frac{1}{\sqrt{2}}    & 0 \cr
0       & 0     & 1 \cr } \right )
\left( \matrix{
1       & 0     & 0 \cr
0       & \frac{1}{\sqrt{3}}    & \frac{2}{\sqrt{6}} \cr
0       & -\frac{2}{\sqrt{6}} & \frac{1}{\sqrt{3}} \cr } \right )
\left( \matrix{
1 & 0 & 0 \cr
0 & 1 & 0 \cr
0 & 0 & 1 \cr } \right ) \; .
%               (16)
\end{eqnarray}
We can read off the following mixing angles:
\begin{equation}
\theta_l = 0 \; , ~~~~
\theta_\nu = \arcsin\frac{1}{\sqrt{2}} =  45^{\circ} \; ,
~~~~ \theta = \arcsin \frac{2}{\sqrt{6}} = 54.7^{\circ} \, .
%               (17)
\end{equation}
We note that $\sin^2 2 \theta_{\nu} = 1$
and $\sin^2  2\theta = 8/9$. Using the arguments given in
Ref. \cite{FX96}, we can also write down the corrections to
the above (lowest-order) 
leptonic mixing matrix, given by the violation of the
democratic symmetry for the charged leptons. 
As an illustrative example, we obtain
\begin{equation}
V'_l = V_l + \frac{m_{\mu}}{m_{\tau}}
\left( \begin{array}{ccc}
0       & \frac{1}{\sqrt{6}}    & -\frac{1}{2\sqrt{3}} \\
0       & \frac{1}{\sqrt{6}}    & -\frac{1}{2\sqrt{3}} \\ 
0       & \frac{1}{\sqrt{6}}    & \frac{1}{\sqrt{3}} 
\end{array} \right) - \sqrt{\frac{m_e}{m_{\mu}}}
	\left( \begin{array}{ccc}
	\frac{1}{\sqrt{6}} & -\frac{1}{\sqrt{2}} & 0 \\
	\frac{1}{\sqrt{6}} & \frac{1}{\sqrt{2}} & 0 \\  
	-\frac{2}{\sqrt{6}} & 0 & 0 \end{array} \right) \, .
%               (18)
\end{equation}
The corrections to $\theta_l$ and $\theta$ are small, i.e., 
$\theta_l = -4.1^{\circ}$ and $\theta = 52.3^{\circ}$. The
value of $\theta_\nu$ is essentially unchanged.
Correspondingly we find
$\sin^2 2 \theta = 0.94$. A similar result for $\sin^2 
2 \theta$ has also been obtained in Ref. \cite{Japan}.

We have obtained large mixing angles for all neutrinos. 
Each flavor eigenstate ($\nu_e$, $\nu_\mu$ or
$\nu_\tau$) is a linear superposition of three mass
eigenstates $\nu_1$, $\nu_2$ and $\nu_3$, described 
by ${V'}^{\dagger}$ instead of $V'$.
The electron neutrino is in lowest order given by
\begin{equation}
\nu_e = \frac{1}{\sqrt{2}} ~ \left (\nu _1 - \nu _2 \right ) \, .
%               (19)
\end{equation}
An electron neutrino produced in the sun would oscillate between the states
$\nu_e = ( \nu_1 - \nu_2 )/\sqrt{2}$ and
$\tilde{\nu}_e = ( \nu_1 + \nu_2 )/\sqrt{2}$.
Note that this state is neither a $\mu$--neutrino nor a $\tau$--neutrino, but
rather a mixture of the two. The solar neutrino experiments are consistent
with a large mixing angle in the scheme of long-wavelength vacuum
oscillations:
\begin{equation}
P(\nu_e \rightarrow \nu_e) \; =\; 1 - \sin^2 2 \theta_{\rm sun}
\sin^2 \left (1.27 \frac{\Delta m^2_{\rm sun} L}{|{\bf P}|} \right ) \;
%               (20)
\end{equation}
with $\sin^2 2 \theta_{\rm sun} \approx 0.7 \ldots 1$ and
$\Delta m^2_{\rm sun} \approx (0.6 \ldots 1.1) \times 10^{-10}
~ {\rm eV}^2$ \cite{Sun}. In our case we have 
$\sin^2 2 \theta_{\rm sun} = \sin^2 2 \theta_{\nu} = 1$. Then
$|m^2_2 - m^2_1 | = \Delta m^2_{\rm sun} \sim 10^{-10} ~ {\rm eV}^2$,
i.e., the two neutrinos $\nu_1$ and $\nu_2$ must be degenerate
to a very high degree of accuracy.

In terms of
mass eigenstates the $\mu$-- and $\tau$--neutrinos are given 
approximately by:
\begin{eqnarray}
\nu_{\mu} & = &\frac{1}{\sqrt{6}} \left( \nu_1 + \nu_2 - 2 \nu_3 \right)
\; , \nonumber \\
\nu_{\tau} & = & \frac{1}{\sqrt{3}} \left( \nu_1 + \nu_2 + \nu_3 \right) \, .
%               (21)
\end{eqnarray}
A $\mu$--neutrino will in general oscillate into all three neutrinos.
However, due to the high degeneracy between the $\nu_1$ and $\nu_2$--states,
oscillations between $\mu$--neutrinos and electron--neutrinos will appear only
at very large distances. Oscillations between $\mu$--neutrinos and
$\tau$--neutrinos could show up at smaller distances, if the mass difference
between the $\left( \nu_1, \nu_2 \right)$--states and the $\nu_3$--state is
sizable. For the
sake of our discussion let us suppose that the first two neutrino states are
completely degenerate, in which case we can perform a 45$^{\circ}$--rotation
among the two states without changing the physical situation.

The atmospheric neutrino
experiments, in particular the recent Superkamiokande measurements 
\cite{SuperK}, are consistent with a large mixing angle in the
$\nu_\mu \leftrightarrow \nu_\tau$ oscillations:
\begin{equation}
P (\nu_\mu \rightarrow \nu_\tau) \; =\; \sin^2 2\theta_{\rm atm}
\sin^2 \left ( 1.27 \frac{\Delta m^2_{\rm atm} L}{|{\bf P}|}
\right ) \;
%               (22)
\end{equation}
with $\sin^2 2\theta_{\rm atm} \approx (0.7 \ldots 1)$ and
$\Delta m^2_{\rm atm} \approx (0.3 \ldots 8) \times 10^{-3}
~ {\rm eV}^2$ \cite{SuperK,Atm}. 
In our case we obtain $\sin^2 2 \theta_{\rm atm}
= \sin^2 2\theta = 0.83$, while $|m^2_3 - m^2_2| =
\Delta m^2_{\rm atm} \sim 10^{-3} ~ {\rm eV}^2$ implies the
weaker degeneracy between $\mu$- and $\tau$-neutrinos.

Thus the observational hints towards neutrino oscillations both for solar and
atmospheric neutrinos indicate a mass pattern for the three neutrino states as
follows. The first two neutrinos $\nu_1$ and $\nu_2$ are almost degenerate,
while the mass of the third neutrino $\nu_3$ is slightly larger or smaller.
If this picture is correct, it would imply that there is
no way to obtain evidence for neutrino oscillations using neutrino beams from
accelerators, unless one performs a long distance experiment. 
For example,
there would be no way to accommodate the results obtained in the 
LSND experiment \cite{LSND}.

The pattern for the neutrino
mass differences discussed above can be realized in two
qualitatively different ways. Either there is a hierarchical pattern in the
neutrino mass spectrum or the three neutrino masses are nearly degenerate. If
the spectrum is hierarchical, the first two mass eigenstates $\nu_{1, 2}$ must
be extremely light. For example one could have 
$m_1 \sim 0$ and $m_2 \sim 10^{-5} ~{\rm eV}$. The third
neutrino $\nu_3$ would have a mass in the range $(0.02 \ldots 0.1)$ eV. Thus
one has $\Delta m^2_{\rm sun} \approx m^2_2$ and
$\Delta m^2_{\rm atm} \approx m_3^2$.
In this case the neutrinos would play only a minor role in
cosmology.

In the case of a degenerate neutrino
mass spectrum the three mass eigenstates would
have nearly the same mass:
\begin{eqnarray}
m_1 & = & \delta_\nu \; , \nonumber \\
m_2 & = & \delta_\nu + \kappa_\nu \; , \qquad (|\kappa_\nu / \delta_\nu
| \ll 1) \; , \nonumber \\
m_2 & = & \delta_\nu + \kappa'_\nu \; , \qquad (|\kappa'_\nu / 
\delta_\nu| \ll 1) \; , 
%               (23)
\end{eqnarray}
with the constraints
$\Delta m^2_{\rm sun} \approx 2 \delta_\nu \kappa_\nu 
\sim 10^{-10} ~ {\rm eV}^2$ and
$\Delta m^2_{\rm atm} \approx 2 \delta_\nu \kappa'_\nu \sim
10^{-3} ~ {\rm eV}^2$.
If neutrinos are a significant part of the dark matter component in the
universe, the sum of the three neutrino states is expected to be in the range
between $4$ eV and
$8$ eV. As an illustrative example we could take $\delta_\nu = 2$ eV,
i.e., $m_1 = 2$ eV, $m_2 = (2 + 2 \times 10^{-11} )$ eV, and
$m_3 = (2 + 10^{-3})$ eV.

I should like to emphasize that we have arrived at these conclusions and
especially at the high degree of degeneracy in the neutrino mass spectrum by
using the constraints from the solar neutrino and atmospheric neutrino
experiments. From the
theoretical point of view discussed initially there would be no need to have
the three neutrino masses to be highly degenerate. It may well be that the
neutrino mass
degeneracy is a hint towards another symmetry (see, e.g., Ref. \cite{VM}
for a SO(10) grand unification model accommodating degenerate neutrino
masses).

Thus far we did not discuss whether the neutrino mass eigenstates are of
Majorana type or Fermi--Dirac type. In fact, both cases are possible. In the
Majorana case one needs to worry about the implications for the neutrinoless
double--$\beta$--decay \cite{Beta}. 
These were discussed in Ref. \cite{FX96}.

Finally we should like to stress again that the case of nearly degenerate
neutrinos, possibly of masses of the order of $2$ eV, is an interesting 
possibility. Large mixing angles are readily obtained in such a model. Both
the mass spectrum and the flavor mixing pattern of the leptons might
differ substantially from those observed for the quarks.

\section{Acknowledgements}
One of us (H.F.) would like to thank Dr.
G. Raffelt and Prof. L. Okun for discussions.

%\begin{figure}
%\vspace*{13pt}
%\leftline{\hfill\vbox{\hrule width 5cm height0.001pt}\hfill}
%\vspace*{1.4truein}             %ORIGINAL SIZE=1.6TRUEIN x 100% - 0.2TRUEIN
%\leftline{\hfill\vbox{\hrule width 5cm height0.001pt}\hfill}
%\fcaption{Radiative Processes for the CP Eigenstates.}
%\label{fig:radk}
%\end{figure}

%\begin{table}[h]
%\tcaption{$\Gamma(K\rightarrow\pi\pi\gamma)$ for the $K^0_S$,
%$K^0_L$ and $K^-$ mesons.}\label{tab:exp}
%\small
%\begin{tabular}{||c|c|c|l||}\hline\hline
%{} &{} &{} &{}\\
%Meson &$\Gamma(\pi^+\pi^-)\; s^{-1}$ &$\Gamma(\pi^+\pi^-\gamma)\; s^{-1}$ &{}\\
%{} &{} &{} &{}\\
%\hline
%{} &{} &{} &{}\\
%$K^0_S$ &$0.769\times 10^{10}$ &$5.46\times 10^7$ 
%&\begin{minipage}{2.5in}
%No DE observed, nor (IB)-E1 interference, despite large
%statistics, for $E^{\ast}_{\gamma}>20 MeV$.
%\end{minipage}\\
%{} &{} &{} &{}\\
%\hline
%{} &{} &{} &{}\\
%\raise13pt\hbox{$K^0_L$} &\raise13pt\hbox{$3.93\times 10^4$} 
%&\raise13pt\hbox{$0.90\times 10^3$}
%&\begin{minipage}{2.5in}
%DE prominent, exceeding IB over the range of measurement
%$20<E^{\ast}_{\gamma}<160 MeV$.
%\end{minipage}\\ 
%{} &{} &{} &{}\\[-37pt]
%{} &{} &(DE $=0.62\times 10^3)$ &{}\\[24pt]
%\hline\hline
%\end{tabular} 
%\end{table}


\section{References}
\begin{thebibliography}{99}
\bibitem{FH} H. Fritzsch and D. Holtmannsp$\rm\ddot{o}$tter,
{\it Phys. Lett.} {\bf B338} (1994) 290.

\bibitem{Dem} See, e.g., H. Harari, H. Haut, and J. Weyers,
{\it Phys. Lett.} {\bf B78} (1978) 459;
Y. Chikashige, G. Gelmini, R.P. Peccei and M. Roncadelli,
{\it Phys. Lett.} {\bf B94} (1980) 499;
Y. Koide, {\it Phys. Rev.} {\bf D28} (1983) 252;
H. Fritzsch, in {\it Proceedings of the Europhysics Topical 
Conference on Flavor Mixing in Weak Interactions}, Erice,
edited by L.L. Chau (Plenum, New York, 1984), p. 717;
C. Jarlskog, in {\it Proceedings of the International Symposium on
Production and Decay of Heavy Hadrons}, Heidelberg 1986, edited
by K.R. Schubert and R. Waldi (DESY, Hamburg), p. 331;
P. Kaus and S. Meshkov, {\it Phys. Rev.} {\bf D42} (1990) 1863;
H. Fritzsch and J. Plankl, {\it Phys. Lett.} {\bf B237} (1990) 451.

\bibitem{FX98} H. Fritzsch and Z.Z. Xing, {\it Phys. Rev.} {\bf D57},
(1998) 594.

\bibitem{PDG} Particle Data Group, R.M. Barnett {\it et al.}, 
{\it Phys. Rev.} {\bf D54} (1996) 1.

\bibitem{FX97} H. Fritzsch and Z.Z. Xing, {\it Phys. Lett.} {\bf B413}
(1997) 396.

\bibitem{Euler} L. Euler, in {\it Memoires de l'Academie de
Sciences de Berlin} {\bf 16} (1760) 176; part of the English
translation can be found in: R. Dugas, {\it A History of
Mechanics} (Dover Publications, 1988) p. 276.

\bibitem{F78} H. Fritzsch, {\it Phys. Lett.} {\bf B73} (1978) 317;
{\it Nucl. Phys.} {\bf B155} (1979) 189.

\bibitem{FX99} H. Fritzsch, Report No. hep-ph/9710409; 
H. Fritzsch and Z.Z. Xing, work in progress.

\bibitem{Stocchi} F. Parodi, P. Roudeau, and A. Stocchi,
Report No. hep-ph/9802289.

\bibitem{FX95} H. Fritzsch and Z.Z. Xing, {\it Phys. Lett.} {\bf B353}
(1995) 114.

\bibitem{Wu} H. Lehmann, C. Newton, and T.T. Wu, 
{\it Phys. Lett.} {\bf B384} (1996) 249.

\bibitem{FX96} H. Fritzsch and Z.Z. Xing, {\it Phys. Lett.} {\bf B372}
(1996) 265.

\bibitem{Japan} M. Fukugita, M. Tanimoto, and T. Yanagida,
{\it Phys. Rev.} {\bf D57} (1998) 4429.

\bibitem{Sun} See, e.g., E. Kearns, talk given at the ITP
Conference on Solar Neutrinos: {\it News About SNUs},
Santa Barbara, December 1997;
G.L. Fogli, E. Lisi, and D. Montanino, Report No. hep-ph/9709473;
N. Hata and P.G. Langacker, {\it Phys. Rev.} {\bf D56} (1997) 6107.

\bibitem{SuperK} Y. Fukuda {\it et al.}, Report No. hep-ex/9803006,
hep-ex/9805006; 
Y. Totsuka, talk given at the 18th International Symposium on
Lepton-Photon Interactions, Hamburg, July 1997.

\bibitem{Atm} M.C. Gonzalez-Garcia {\it et al.},
Report No. hep-ph/9712238; 
V. Barger, T.J. Weiler, and K. Whisnant, Report No. hep-ph/9712495.

\bibitem{LSND} C. Athanassopoulos {\it et al.}, {\it Phys. Rev. Lett.}
{\bf 77} (1996) 3082.

\bibitem{VM} A. Ioannissyan and J.W.F. Valle, {\it Phys. Lett.}
{\bf B332} (1994) 93;
D.O. Caldwell and R.N. Mohapatra, {\it Phys. Rev.} {\bf D48}
(1993) 3259.

\bibitem{Beta} See, e.g., H.V. Klapdor-Kleingrothaus, Report No.
hep-ex/9802007; and references therein.
\end{thebibliography}
\end{document}